\documentclass[aps,prd,reprint,groupedaddress,preprintnumbers,nofootinbib]{revtex4-2}
\usepackage{amsfonts,bm,amssymb,euscript,array,babel}
\usepackage{mathtools}
\usepackage{tikz-cd}
\usepackage{amsmath}
\usepackage{hhline}
\usepackage[amsmath]{empheq}
\usepackage{hyperref}
\usepackage{cancel}
\usepackage{mathrsfs}
\usepackage{lipsum}
\usepackage{adjustbox}

\newcommand{\dd}{\mathrm{d}}

\newcommand{\be}{\begin{equation}}
\newcommand{\ee}{\end{equation}}

\def \bea{\begin{eqnarray}} 
\def\eea{\end{eqnarray}}
\def\bse{\begin{subequations}}	
	\def\ese{\end{subequations}}
\def\bal{\begin{align}} 
\def\eal{\end{align}}

\def\mc{\mathcal}

\def\bi{\begin{itemize}} 
	\def\ei{\end{itemize}}

\def\a{\alpha} \def\b{\beta} \def\g{\gamma}  \def\d{\delta} 
 
  \def\h{\eta} 
\def\l{\lambda}  \def\m{\mu}
\def\n{\nu}    
\def\s{\sigma}   
  
\def\O{\Omega}

\begin{document}

%\preprint{RBI-ThPhys-2020-25}

\title{\large Torsion-induced gravitational $\theta$ term and gravitoelectromagnetism}

\author{Athanasios Chatzistavrakidis}
\email{Athanasios.Chatzistavrakidis@irb.hr}
\author{Georgios Karagiannis}
\email{Georgios.Karagiannis@irb.hr}
%\homepage[]{Your web page}
%\thanks{}
%\altaffiliation{}
\affiliation{Division of Theoretical Physics, Rudjer Bo\v skovi\'c Institute, Bijeni\v cka 54, 10000 Zagreb, Croatia}
\author{Peter Schupp}
\email{p.schupp@jacobs-university.de}
\affiliation{Department of Physics and Earth Sciences, Jacobs University, 28759 Bremen, Germany }

\begin{abstract}
Motivated by the analogy between a weak field expansion of general relativity and Maxwell's laws of electrodynamics, we explore  physical consequences of a parity violating $\theta$ term in gravitoelectromagnetism. This is distinct from the common gravitational $\theta$ term formed as a square of the Riemann tensor. Instead it appears as a product of the gravitoelectric and gravitomagnetic fields in the Lagrangian, similar to the Maxwellian $\theta$ term. We show that this sector can arise from a quadratic torsion term in nonlinear gravity. 
 In analogy to the physics of topological insulators, the torsion-induced $\theta$ parameter can lead to excess mass density at the interface of regions where $\theta$ varies and consequently it generates a correction to Newton's law of gravity. We discuss also an analogue of the Witten effect for gravitational dyons. 
\end{abstract}

\maketitle

\section{Introduction}

It is well known that in gauge theories, such as electrodynamics and Yang-Mills, there exists a Lorentz and gauge invariant, CP-violating, topological $\theta$ term, which is quadratic in the field strength. 
In covariant notation, the abelian electromagnetic $\theta$ term is 
\be \label{thetaMaxwell}
{\cal S}_{\theta}^{\text{EM}}=\frac{ e^2}{64\pi^2}\int \dd^4 x\,\theta\,\varepsilon^{\m\n\rho\sigma}F_{\m\n}F_{\rho\sigma}\,,
\ee 
with its non-abelian counterpart involving also a trace over the gauge indices of the non-Abelian gauge group. In terms of the electric and magnetic field components of the Faraday tensor $F_{\m\n}$, this term is proportional to $\vec E\cdot \vec B$. Moreover, it can be identified with the derivative of the three-dimensional Chern-Simons form. 
 Despite being a total derivative, the $\theta$ term plays an important role in several physical situations. A number of  applications are collected and explained in Refs. \cite{FoF,Tong,AlvarezGaume:1996mv}. 

On the other hand, there exist regimes where gravity and electromagnetism formally appear very similar. For instance, this is true for weak gravitational fields generated by rotating masses. A celebrated example is the Lense-Thirring effect, which has important astrophysical implications 
and whose experimental verification has been a main objective of Gravity Probe~B~\cite{Everitt:2011hp}. 
In such situations, an accelerating mass can produce a gravitomagnetic field in much the same way as an accelerating charge produces a magnetic field. This is a prediction of general relativity, which in cases relevant for such effects may be brought to a form that resembles Maxwell's equations. This formulation is referred to as gravitoelectromagnetism (GEM)~\cite{Mashhoon:2003ax}. 

In this letter, we suggest that the above two statements may be combined. We propose a new gravitational $\theta$~term that plays the role of \eqref{thetaMaxwell} in GEM and discuss some of its physical consequences. One's first thought when considering such a term in nonlinear gravity would be a quadratic invariant in the curvature tensor, the Pontryagin density, which is moreover the derivative of the gravitational Chern-Simons term. This is structurally similar to the Maxwellian $\theta$ term and it has the form 
\be \label{RR}
{\cal S}^{\text{GR}}_{\theta}=\frac {\a}{8}\int \dd^4 x\,\theta\,\varepsilon^{\m\n\rho\sigma}R^{\kappa}{}_{\l\m\n}R^{\l}{}_{\kappa\rho\sigma}\,,
\ee 
in terms of the Riemann tensor. Here, $\a$ is a dimensionful parameter, which may be identified with $(16\pi G)^{-1}$, as for example in Ref. \cite{Jackiw:2003pm}.
 This term has been considered in several contexts, for example in addressing the strong CP problem \cite{Dvali:2005an,Karananas:2018nrj}, explaining neutrino masses \cite{Dvali:2016uhn}, studying transport properties of black hole horizons \cite{Fischler:2016jaq} and exploring phenomenological consequences of Chern-Simons gravity around a massive spinning body \cite{Smith:2007jm}, to name a few. Note also that although general relativity is parity preserving, the question of whether gravity can be parity violating is not settled. For example, this has been addressed in the context of gravitational waves in recent years, as in Refs. \cite{Alexander:2017jmt,Nishizawa:2018srh}. 

In this work, however, it is not the Pontryagin density term \eqref{RR} that we are interested in. Instead, we ask whether there exists an alternative gravitational $\theta$ term, which shares with the Maxwellian one the property that it is of the form $\vec E\cdot\vec B$, with $\vec E$ and $\vec B$ this time being the gravitoelectric and gravitomagnetic fields of GEM. Clearly, this combination does not arise from \eqref{RR}, since the decomposition of the Riemann, respectively Weyl, tensor into ``electric'' and ``magnetic'' components \cite{Bel,Hawking:1966qi} leads to one additional derivative compared to each of the GEM fields. 
 From the viewpoint of nonlinear gravity, we show that such a combination may appear in the weak field limit from a quadratic term of the form 
\be \label{theta GEM}   
{\cal S}_\text{$\theta$G}=\frac{ c^4}{16\pi G} \int \dd^4 x\,\theta\,\varepsilon^{\m\n\rho\s} \O_{\m\n\l} \O_{\rho\s}{}^\l\,,
\ee 
where $\Omega$s are coefficients of anholonomy, directly related to torsion and of first order in derivatives, just like the Faraday tensor in  \eqref{thetaMaxwell}. 
 It is worth mentioning that the relation of this term with torsion is reasonable in view of the fact that ``parity violation and torsion go hand-in-hand'', as explained in section 21 of Ref. \cite{Freund:1986ws}. This suggests that \eqref{theta GEM} can be relevant in theories of gravity such as Einstein-Cartan or Teleparallel Equivalent General Relativity (TEGR).  
Topological invariants involving the torsion tensor were first constructed in \cite{Nieh:1981ww} and further studied in \cite{Chandia:1997hu,Li:1999ue,Obukhov:1997pz}. They have also been used in the context of a gravitational Peccei-Quinn mechanism \cite{Mercuri:2009zi}, in studies of axions in torsional gravity \cite{Castillo-Felisola:2015ema} and in the context of AdS$_4$/CFT$_3$ holography \cite{Leigh:2008tt}.

 As far as physical applications are concerned, recall that in the abelian case \eqref{thetaMaxwell}, interesting phenomena arise when $\theta$ does not take the same value everywhere. For example, topological insulators in condensed matter physics are materials characterized by the fact that there is an interface where $\theta$ changes value. Within the material $\theta=\pi$, whereas everywhere else $\theta$ takes a vanishing vacuum value. 
These are precisely the two possible values of~$\theta$ for which CP, or equivalently time-reversal symmetry, is not violated. In a similar fashion, if we make the assumption that the gravitoelectromagnetic $\theta$ parameter arising in the weak field limit of \eqref{theta GEM} is a non-constant fixed background, we derive a modified set of GEM equations. Such a setup would have profound physical consequences: In the gravitational analogs of topological magnetoelectric effects, a gravitoelectric (Newtonian) field can induce gravitomagnetic forces (with frame dragging) and vice versa. 

In this letter we investigate two examples.
The first is 
a gravitational analogue of topological magnetoelectric effects, similar in spirit to the Hall effect or topological insulators as mentioned above, where excess mass density or mass current appear at the interface of regions with different $\theta$. We demonstrate that this can provide a correction to the Newtonian gravitational field in physical settings where a gravitomagnetic field is present. The second application is the analogue of the Witten effect for dyons. In this case, we consider the gravitational analogue of pointlike magnetic monopoles, coined ``gravitipoles''  by Zee in~\cite{Zee:1986sr}, showing that in a region with nonvanishing $\theta$ they appear to have mass aside their gravitomagnetic charge. Such gravitipoles could arise as mirror images of ordinary massive objects in the presence of the GEM equivalent of topological surface states~\cite{2009Sci...323.1184Q}.
              
\section{Torsion-induced $\theta$ term in gravity}

The standard formulation of General Relativity features an Einstein-Hilbert action that looks rather different from the action of gauge theories: It is not quadratic in field strength tensors, but rather linear in the Riemann tensor, which itself is written in terms of non-tensorial Christoffel symbols. This aesthetic deficit can be overcome with the help of a local orthonormal frame 
$e^a = e_\mu^a(x) \,\dd x^\mu$ and its dual $e^\mu_a(x)$, with anholonomy coefficients $\O_{\m\n}{}^\s = \partial_{[\m} e_{\n]}{}^a\, e_a{}^\s$: A combination of the three parity invariant Weitzenb\"ock invariants~\footnote{Recall that Weitzenb{\"o}ck invariants transform as densities under general coordinate transformations, as scalars under Lorentz transformations and they are quadratic in first derivatives of the vielbeins \cite{Ortin:2015hya}.},
gives an alternative gravity action quadratic in the anholonomy coefficients 
\be \label{TGaction}\begin{split}
{\cal S}_\text{TG} = \frac{c^4}{16\pi G}\int \dd^4 x \sqrt{|g|}\,(\O_{\n\m\rho} &\O^{\n\m\rho} + 2 \O_{\n\m\rho} \O^{\rho\m\n} \\
&- 4 \O_{\l\m}{}^\m \O^\l{}^\n{}_\n) \,.
\end{split}
\ee 
Realizing that $T_{\m\n}{}^\rho = -2\O_{\m\n}{}^\rho$ is the torsion of the curvature-free but torsion-full Weitzenb\"ock connection, this is in fact the starting point of TEGR. In terms of the contorsion tensor, given by 
\be 
K_{\m\n\rho} = \O_{\m\n\rho} - \O_{\n\rho\m} + \O_{\rho\m\n}\,,
\ee 
the action reads \cite{Ortin:2015hya,Hayashi:1979qx}
\be
{\cal S}_\text{TG} = \frac{c^4}{16\pi G}\int \hspace{-0.1cm}\dd^4 x \sqrt{|g|}\left(K_{\n\l} {}^{\n}K_\mu{}^{\m\l}-K_{\m\l}{}^\n K_\n{}^{\m\l}\right)
\ee
and using the identity 
\be  \label{RKidentity}
\frac12 R^\text{LC}_{\m\rho\s\n} = K_{[\m|\n\l} K_{|\rho]\s}{}^\l
- \nabla_{[\m} K_{\rho]\s\n}  \,,
\ee
it is found to be equal to the Einstein-Hilbert action~${\cal S}_\text{EH}$ plus the total derivative term $2\nabla_\mu K_\nu{}^{\nu\mu} = 4 \nabla_\mu \Omega^{\mu\nu}{}_\nu$. Therefore, the two formulations are classically equivalent.
${\cal S}_\text{EH}$ depends on the frame fields $e_\mu^a$ only via $g_{\mu\nu} = e_\mu^a e_\nu^b \eta_{ab}$ and is thus manifestly invariant under local Lorentz transformations.

Now that we have rewritten the gravity action in gauge-theory akin form, we can start looking for a gravitational analogue of the famous gauge theory $\theta$ term.  The natural choice that is topological and quadratic in the contorsion tensors (or holonomy coefficients) is
 \footnote{A fourth, parity-violating Weitzenb{\"o}ck invariant given as $\varepsilon^{\m\n\rho\sigma}\Omega_{\m\l}{}^{\l}\Omega_{\n\rho\sigma}$ in 4 dimensions \cite{Ortin:2015hya}, which can be extended to any dimension, is not identical to the parity-violating term~(\ref{theta TEGR}).}
\be\label{theta TEGR}
\begin{split}
{\cal S}_\text{$\theta$G} &= \frac{c^4}{16\pi G}\int \dd^4 x \, \theta \, \varepsilon^{\m\n\rho\s} K_{[\m\n]\l} K_{[\rho\s]}{}^\l \\
&= \frac{c^4}{16\pi G}\int \dd^4 x \, \theta \, \varepsilon^{\m\n\rho\s} \O_{\m\n\l} \O_{\rho\s}{}^\l   \,.
\end{split}
\ee
Contracting \eqref{RKidentity} with the Levi-Civita symbol $\varepsilon$ gives zero 
and implies that \eqref{theta TEGR} is a total derivative for constant $\theta$: $\varepsilon^{\mu\nu\rho\sigma} K_{\mu\nu\lambda} K_{\rho\sigma}{}^\lambda
= \sqrt{|g|} \nabla_\mu (\epsilon^{\mu\nu\rho\sigma}K_{\rho\sigma\nu})
= \partial_\mu (\varepsilon^{\mu\nu\rho\sigma} K_{\rho\sigma\nu})$, where $\epsilon$ is the Levi-Civita \emph{tensor}. Note that this is true only for the Weitzenb\"ock connection. A more general identity holds for arbitrary connections with curvature \cite{Nieh:1981ww}, giving rise to the Nieh-Yan topological invariants \cite{Chandia:1997hu,Li:1999ue}. 
Rewritten in terms of frame fields 
\be
{\cal S}_\text{$\theta$G} %&= \frac{c^4}{16\pi G}\int \dd^4 x \, \theta \, \varepsilon^{\m\n\rho\s} \eta_{ab}\partial_\mu e_\nu^a \partial_\rho e_\sigma^b  \\ 
=  \frac{c^4}{64\pi G}\int \theta\, \eta_{ab} \,\dd e^a \wedge \dd e^b \,,
\ee
which is strikingly similar to the gauge theory $\theta$ term $\theta\, \dd A \wedge \dd A$ with $F = \dd A$ in \eqref{thetaMaxwell} and  has the gauge symmetry $\delta e^a = \dd \lambda^a$. 

As in the gauge theory case, one can make the hypothesis that there may be regions in space with different values of $\theta$ and in the transition regions one needs to consider a non-constant background field $\theta(t,\vec x)$. We complete the model by adding matter,
\be {\cal S} = {\cal S}_\text{G} + {\cal S}_\text{$\theta$G} + {\cal S}_\text{M}\,,
\ee
where ${\cal S}_{\text{M}}$ is the matter action whose variation with respect to the metric yields the energy momentum tensor and ${\cal S}_\text{G}$ can be any gravity action that can be (re-)written in terms of frame fields: ${\cal S}_\text{TG}$, ${\cal S}_\text{EH}$, or Einstein-Cartan. 

In the linearized limit, i.e. upon considering linear perturbations of the metric, we have $g_{\mu\nu} \simeq \eta_{\mu\nu} + h_{\m\n}$ and 
$e_\m{}^a \delta_{a\n} \simeq \delta_{\m\n} +\frac12 (h_{\m\n} + b_{\m\n})$, where $b_{\m\n}$ is the antisymmetric Kalb-Ramond field.
The action $S_\text{TG}$ becomes the usual linearized gravity action (massless Fierz-Pauli theory) and the Kalb-Ramond 2-form $b_{\m\n}$ decouples due to the special form of \eqref{TGaction} -- we shall ignore it in the following. Hence, $2K_{[\m\n]\rho}=2\O_{\m\n\rho} \simeq  \partial_{[\mu} h_{\nu]\rho} \simeq -K_{\rho\m\n}$.
Introducing the trace-reversed metric $\bar h_{\m\n}:=h_{\m\n}-\frac{1}{2}\,\h_{\m\n}\,h$ and imposing de Donder gauge $\partial^\mu \bar h_{\mu\nu}\overset{!}{=}0$, the action further simplifies to
\be \label{theta tegr linearized}\begin{split}
{\cal S} \simeq \frac{c^4}{64\pi G}\int \dd^4x\,(&h_{\m\n}\,\Box \,\bar h^{\m\n}\\
&+\theta\,\varepsilon^{\s\rho\l\m}\partial_{\rho}h_{\m\n}\,\partial_\l \bar h_\s^\n)+\mathcal{S}_{\text{M}}\,.
\end{split}
\ee
For constant $\theta$, the second term is topological and does not contribute to the dynamics. In general, however, the action \eqref{theta tegr linearized} yields the field equations
\be \label{eoms theta tegr}
\Box\,\bar h_{\m\n}-\partial^\a\theta\,\varepsilon_{\a\b\g(\m}\,\partial^\b \bar h^{\gamma}_{\n)}=-\frac{16\pi G}{c^4}\,T_{\m\n}\,,
\ee
where $T_{\m\n}$ is the conserved energy-momentum tensor that results from variation of the matter sector $\mathcal{S}_{\text{M}}$.
To analyze these equations, one has to make some assumptions about the matter content of the theory: For a finite distribution of dust, the energy-momentum tensor has the form  $T_{\mu\nu}=\rho\,v_\mu v_\nu$, where $\rho$ is the mass density. If the dust is moving slowly, i.e. $|\Vec{v}|\ll c$, then the components of $T_{\m\n}$ are $T_{00}\simeq\rho \,c^2$,  $T_{i0}\simeq-\rho \,v_i\,c$ and $T_{ij}\simeq\rho \,v_i\,v_j$. Consequently, the components of any solution of equation \eqref{eoms theta tegr} will scale as $\bar h_{00}=\mc{O}(c^{-2})$,  $\bar h_{i0}=\mc{O}(c^{-3})$ and $\bar h_{ij}=\mc{O}(c^{-4})$.  In the ensuing we will ignore all terms and effects that are of order $\mc{O}(c^{-4})$, including the purely spacelike components $\bar h_{ij}$.

Finally, in this work we are going to use the mostly-plus signature convention for the Minkowski metric.

\section{\label{sec3}Axion Gravitodynamics}

The set of GEM equations is obtained in terms of gravitoelectic and gravitomagnetic potentials, which are defined as 
\be 
\phi:=-\frac{c^2}{4}\bar{h}_{00}\quad  \text{and} \quad A_{i}:=\frac {c^2}{2}\bar{h}_{i0}\,.
\ee 
Here, the gravitoelectric potential $\phi$ is identified with the Newton potential. 
%{\mc A}^\m \equiv (\phi,\sfrac 12\vec { A}) := -\frac{c^2}4 \bar h^{\mu 0}$. 
Both potentials have dimensions of energy over mass and they scale with the speed of light as $\phi=\mc{O}(c^0)$ and $\vec{ A}=\mc{O}(c^{-1})$. The temporal component of the de Donder gauge is equivalent to the analogue of the Lorentz gauge 
$
\frac 2c \frac{\partial\phi}{\partial t}+\vec{\nabla}\cdot \vec{A} = 0\,,
$
 and
the spatial components  give $\frac{1}{c^3}\,\frac{\partial{\vec{ A}}}{\partial t}=\vec{0}$.
In addition, one can define the gravitoelectric and gravitomagnetic fields
by 
%$ F_{\m\n} = \partial_\m {\mc A}_\n - \partial_\n \mc A_\m$, i.e. 
$\vec{{E}}:=-\vec{\nabla}\phi-\frac{1}{2c}\frac{\partial{\vec{A}}}{\partial t}\,,\quad \vec{ B}:=\vec{\nabla}\times\vec{ A}$,
which both have dimensions of acceleration and scale with the speed of light as $\vec{ E}=\mc{O}(c^0)$ and $\Vec{ B}=\mc{O}(c^{-1})$. Clearly, the second term in the definition of $\vec{E}$ does not contribute at lowest order. Combining the Lorentz gauge condition %\eqref{de donder} 
with the equations of motion~\eqref{eoms theta tegr}, one can easily see that the GEM fields satisfy the equations
\begin{subequations}
\begin{align}
\label{Maxwell 1}&\vec{\nabla}\cdot\vec{ E}=-4\pi G\rho-\frac 12\vec{\nabla}\theta\cdot\vec{ B}\,,\\
\label{Maxwell 2}&\vec{\nabla}\cdot\vec{ B}=0\,,\\
\label{Maxwell 3}&\vec{\nabla}\times\vec{ E}+\frac{1}{2c}\,\frac{\partial{\vec{ B}}}{\partial t}=\vec{0}\,,\\
\label{Maxwell 4}&\vec{\nabla}\times\vec{ B}-\frac{2}{c}\,\frac{\partial{\vec{ E}}}{\partial t}=-\frac{8\pi G}{c}\,\rho \,\vec{v}+\vec{\nabla}\theta\times\vec{ E}+\frac{1}{2c}\,\frac{\partial\theta}{\partial t}\,\vec{ B}\,.
\end{align}\label{Maxwell}
\end{subequations}
These equations are formally analogous to Maxwell's equations in the presence of non-constant $\theta$, as in the invisible axion models \cite{Sikivie:1983ip} and in axion electrodynamics \cite{Wilczek:1987mv}. Hence, the gravitomagnetic field $\vec{ B}$ induces mass density in regions of space where $\theta$ varies, as evident from the analogue of Gauss' law \eqref{Maxwell 1}, and mass current whenever $\theta$ varies in time, as evident from the analogue of Amp\`ere's law \eqref{Maxwell 4}. In the first case, namely for spatial variation of $\theta$, mass current is also effectively induced by the gravitoelectric field $\vec{ E}$.
Therefore, varying $\theta$ may have interesting physical consequences, two of which we outline in the next section. 

As a final remark, one can notice that the GEM equations \eqref{Maxwell} are obtained as Euler-Lagrange equations from the action 
\be 
{\cal S}_{\theta\text{-GEM}}\simeq \frac{1}{8\pi G}\int \dd^4x\left(E^2-B^2-2\theta\vec{E}\cdot\vec{B}\right)+\mathcal{S}_{\text{M}}
\ee
which is nothing but \eqref{theta tegr linearized} in terms of GEM fields, the last term being the matter action. As expected, the theta term is obtained as the dot product of the gravitoelectric and gravitomagnetic fields, much like its counterpart in classical electrodynamics.

\section{Some physical consequences} 

\subsection{Correction to Newton's law} % due to gravitomagnetic field}
As we already mentioned, one consequence of \eqref{Maxwell} is that a gravitomagnetic field in a region of space with nonvanishing and constant $\theta$ induces a gravitoelectric field outside this region where $\theta=0$. 
This is in full analogy to what happens when one considers a topological insulator having $\theta=\pi$. In that case, a magnetic field perpendicular to the insulator's surface induces an effective surface charge density. In turn, this produces an electric field in the exterior region where $\theta=0$.

For example, this is the case when we consider a uniform spherical mass distribution of radius $R$ and $\theta\ne 0$ having a rotating massive body of mass $M$ and angular momentum $\vec{S}$ at its center. In addition, we require that the radius $R$ is much larger than the characteristic scale of the body. This will ensure that the GEM fields generated by the rotating body will have their standard far-field form \cite{Clark:2000ff}
\be 
\Vec{E}(\vec r)\simeq -G M\frac{\hat r}{r^2}\,,\quad \vec{B}(\vec r)\simeq \frac{G}{cr^3}\left[\vec S-3(\vec S\cdot\hat r)\,\hat r\right]
\ee
in the internal vicinity of the spherical boundary.
Assuming that $\vec{S}$ points to the north and denoting the polar angle by $\vartheta$, one gets the relation $\vec S\cdot\hat r=S\cos\vartheta$. It is also obvious that the gradient of $\theta$ reads as $\vec{\nabla}\theta( r)=-\theta\,\delta(R-r)\,\hat r$. Then Eq. \eqref{Maxwell 1} implies that there is an effective accumulation of mass density on the spherical shell,
\be \label{eff density}
\rho_\text{eff}(r,\vartheta)=
\frac{1}{8\pi G}\,\vec{\nabla}\theta(r)\cdot\vec{B}(r,\vartheta) =
\frac{S\theta}{4\pi c}\,\frac{\d(r-R)}{r^3}\,\cos\vartheta\,.
\ee
As expected, the mass density \eqref{eff density} induces an effective gravitoelectric field outside the distribution. To calculate this, one needs to consider a spherical shell that is homocentric to the mass distribution but has radius $r>R$. This will serve as the Gaussian surface. Integrating Eq. \eqref{Maxwell 1} over the region enclosed by this surface, leads to the gravitoelectric field
\be 
\vec{ E}(r,\vartheta)\simeq- G\left(M-\frac{S\theta}{cR}\,\cos\vartheta\right)\frac{\hat r}{r^2}\,,\quad \forall\, r>R\,.
\ee
Thus, we have found that the nonvanishing $\theta$ parameter induces a correction to the gravitoelectric field produced by the actual mass of the source and therefore to the Newtonian gravitational potential. 

It is worth mentioning that a similar analysis for spinning spherical bodies in the presence of the term \eqref{RR} appears in Ref. \cite{Smith:2007jm}. This differs from the present one in two ways: Firstly, the modifications to the GEM equations due to the Chern-Simons term are different, in particular Amp\`ere's law becomes a second-order differential equation for $\vec{B}$. Secondly, the corresponding $\theta$ parameter is assumed to depend only on time and, thus, leads to a modification to the gravitomagnetic field through Amp\`ere's law.

\subsection{Gravitational Witten effect} 

In a different spirit, one can follow \cite{Zee:1986sr} and speculate the existence of gravitipoles. These objects are the gravitational analogues of pointlike magnetic monopoles, for which Eq. \eqref{Maxwell 2} is modified to
\be 
\vec{\nabla}\cdot\vec{ B}=4\pi G\,\widetilde M\,\delta^{(3)}(\vec r)\,,\quad \widetilde M>0\,.
\ee
Note that the gravitomagnetic charge $\widetilde M$ also has dimensions of mass. One of the physical consequences of such an assumption is that planetary orbits would be displaced had there been a gravitipole in the sun \cite{Zee:1986sr}, to the extend that the sun would not lie at the centre of the orbits. In addition, Zee argued that the existence of gravitipoles would lead to mass quantization. Although these ideas are certainly speculative, let us entertain the possibility of existence of gravitipoles and examine what happens when it is combined with the $\theta$ term introduced in the previous sections. 

Recall that in gauge theory, for example in the Georgi-Glashow model, the presence of a $\theta$ term leads to the effect that a magnetic monopole always carries an electric charge, thus being essentially a dyon \cite{Witten:1979ey}. We now ask the question whether a similar conclusion holds for gravitipoles{\footnote{The Witten effect in the context of the Poincar\'e gauge theory of gravity with the term (\ref{RR}) was studied in \cite{Mielke:1985kz}.}}.   
To answer that, referring to the $\theta$ parameter of GEM, let us assume that $\theta\ne 0$ everywhere apart from a hollow sphere of radius $R$, where $\theta=0$. The gradient of $\theta$ is, therefore, $\vec{\nabla}\theta(r)=\theta\,\d(r-R)\,\hat r$. Placing now a gravitipole at the center of the hollow sphere induces a radial gravitomagnetic field  
\be 
\vec{ B}(r)=G\widetilde M\,\frac{\hat r}{r^2}\,.
\ee
This field induces a mass density at the interface, of the form
\be 
\rho_{eff}(r)
=\frac{1}{8\pi G}\,\vec{\nabla}\theta(r)\cdot\vec{ B}(r)
=\frac{\theta\widetilde M}{8\pi}\,\frac{\d(r-R)}{r^2}\,,
\ee
which in turn
generates an effective gravitoelectric field
outside the sphere. To calculate this, one needs to choose a spherical shell that is homocentric to the hollow sphere but has radius $r>R$ as Gaussian surface. Doing so, one easily finds that $\vec{E}_{eff}(r)=-\frac{\theta}{2}\vec B(r)$. 

The crucial point here is that neither field depends on the radius $R$. In other words, one can take the limit $R\to 0$ and shrink the sphere until it becomes a point---this being true also in the standard case of Witten effect \cite{Tong}. Then, all of space has $\theta\ne 0$ and we observe that the gravitipole acquires mass. This is then a gravitational analogue of the Witten effect for magnetic monopoles, which indicates that a gravitipole becomes a gravitational dyon when placed inside a region of space with $\theta\neq 0$. The mass of the dyon is related to its gravitomagnetic charge by $M=\frac{\theta}{2}\widetilde M$. 

We conclude this section with a final remark: In analogy to magnetic monopoles, the radial gravitomagnetic field can be locally expressed as $\vec { B} = \nabla \times \vec{ A}$ in terms of a vector potential $\vec { A}$, which is given up to gauge transformations by its azimuthal component ${ A}_{\varphi} = - G\widetilde M \cos\vartheta$ (w.r.t.\ the coordinate basis $\dd \varphi$). This corresponds to a NUT charge of $N = -\frac{G\widetilde M}{c^2}$, because $g_{\varphi t} = \bar h_{\varphi t} = -\frac{2 G\widetilde M}{c^2}\cos\vartheta = 2 N \cos \vartheta$. 
%(Maybe the NUT charge is $N = -\frac{G\widetilde M}{c^2}$ instead?)

\section{Conclusions \& Outlook} 

Topological $\theta$ terms have a long history and they have found a plethora of physical applications in both high energy and condensed matter physics. In this letter, we suggested a novel $\theta$ term in gravitoelectromagnetism, having its origin in a topological invariant involving the torsion tensor in nonlinear gravity, and we studied some of its physical consequences. This term modifies the field equations of gravitoelectromagnetism and thus leads to a fine interplay of the gravitoelectric (Newtonian) potential and the gravitomagnetic field, in a similar fashion to standard electrodynamics. Specifically, near interfaces where $\theta$ changes value, one is induced by the other. As an example of this, we have demonstrated that the Newtonian gravitational potential can receive a $\theta$-dependent correction in presence of a gravitomagnetic field. It would be interesting to investigate whether
such effects could be relevant in astrophysical contexts, as an
alternative or in synergy with dark matter. Furthermore, it would be also interesting to study the opposite effect, i.e. gravitomagnetic fields induced by static Newtonian ones in presence of nonvanishing $\theta$.

In addition, we studied the corresponding analogue of Witten's effect in presence of the new $\theta$ term. 
Note that apart from high energy physics, this physical effect has also been studied in condensed matter physics, for example in relation to fracton phases \cite{PhysRevB.96.125151}. In the present context, making the assumption of existence of gravitipoles, we showed that the latter acquires a mass in a region of nonvanishing $\theta$, thus turning into a gravitational dyon. 

Finally, another arena where this $\theta$ term can serve as a starting point for further studies is parity violating gravity. It has been argued that signatures of parity violation may appear in observations of gravitational waves \cite{Alexander:2017jmt,Nishizawa:2018srh}, mainly in relation to Chern-Simons modified gravity. This possibility for the torsion-induced parity-violating term considered in this paper 
deserves a careful analysis too. 

\section*{Acknowledgements}
We would like to thank Cristobal Corral for comments on the first version of the paper and for bringing to our attention Refs. \cite{Chandia:1997hu,Mercuri:2009zi,Castillo-Felisola:2015ema}.
This work is supported by the Croatian Science Foundation Project ``New Geometries for Gravity and Spacetime'' (IP-2018-01-7615), and also partially supported by the European Union through the European Regional Development Fund - The Competitiveness and Cohesion Operational Programme (KK.01.1.1.06). We express our gratitude to the DFG Research training Group 1620 ``Models of Gravity''.

% Create the reference section using BibTeX:
\bibliographystyle{ieeetr}
\bibliography{new}

\end{document}